\documentclass[runningheads,a4paper]{llncs}

\usepackage{amssymb}
\usepackage{graphicx}
\usepackage{cite}
\usepackage{url}
\urldef{\mailsa}\path|{ju.zhang, a.tordeux, a.seyfried}@fz-juelich.de|
\newcommand{\keywords}[1]{\par\addvspace\baselineskip
\noindent\keywordname\enspace\ignorespaces#1}

\begin{document}

\mainmatter  

\title{Effects of boundary conditions on single-file pedestrian flow}

\titlerunning{Effects of boundary conditions on single-file pedestrian flow}

%
%
\author{Jun Zhang$^1$, Antoine Tordeux$^1$\and Armin Seyfried$^{1,2}$}
\institute{$^1$ J\"ulich Supercomputing Centre,
 Forschungszentrum J\"ulich GmbH, Germany\\
$^2$ Computer Simulation for Fire Safety and Pedestrian Traffic,
  Bergische Universit\"at Wuppertal, Germany\\
\mailsa\\}

\maketitle

\begin{abstract}
The characteristics of pedestrian flows are strongly influenced by several factors. In this paper we investigate effects of boundary conditions on one dimensional pedestrian flow which involves purely longitudinal interactions. In qualitatively stop-and-go waves were observed under closed boundary condition and dissolve when the boundary is open. To get more detailed information the fundamental diagrams are compared by using Voronoi-based measurement method. Higher maximal specific flow is observed from the pedestrian movement at open boundary conditions.
\end{abstract}

\section{Introduction}

The investigation of pedestrian dynamics is not only of scientific interest, e.g. due to the various collective phenomena that can be observed, but also of great
practical relevance. Understanding and predicting the dynamical properties of larger crowds is of great importance to avoid or reduce the number of casualties under emergencies or to improve the capacity of a pedestrian facility in normal situation \cite{Schadschneider2009, Schadschneider2009a}. Even for the basic fundamental diagram, however, large discrepancies are shown in previous studies \cite{Seyfried2010, Zhang2012} and is influenced by several factors including pedestrian motivation, facility geometry, culture differences etc.

Facing such problems, one dimensional pedestrian flows which involve purely longitudinal interactions and simplify the possible influences on pedestrian have been studied from various aspects in recent years. Seyfried et al. measured the fundamental diagram of single file flow for the density up to 2 $1/m$ \cite{Seyfried2005}. The same experiments were carried out in India by Chattaraj et al. \cite{Chattaraj2009} and in China by Liu et al. \cite{Liu2009} to investigate the culture difference on the fundamental diagram. Jeli\'{c} et al. conducted the similar experiment inside a ring corridor formed by inner and outer circular walls to study the properties of pedestrians moving in line \cite{Jelic2012, Jeli`c2012}. The density in their experiment reaches 3 $1/m$ and the stepping behavior and fundamental diagrams are studied. From this study, three regimes (free regime, weakly constrained regime and strongly constrained regime) are distinguished by analysis the velocity-spatial headway relationship. In this way, the pedestrian flow with higher velocity (free flow state) is divided into two different regimes but the flow for the velocity $v < 0.8~m/s$, which belongs to congested states, is not distinguished. However, the properties of pedestrian flow under high density situation are much more interesting. Stop-and-go wave \cite{Seyfried2010b} is a special phenomenon of congested flow but it appears in only part of ranges of this state. On the other hand, the influences of boundary conditions, open or closed, on the fundamental diagram are still need studying.

In this paper, we present new experimental results with higher density range to study the prosperities of single file pedestrian flow. The fundamental diagrams will be analyzed and the influence of different boundary conditions on it will be studied.

\section{Experiment Setup}

\begin{figure}
\centering{
\includegraphics[height=6 cm]{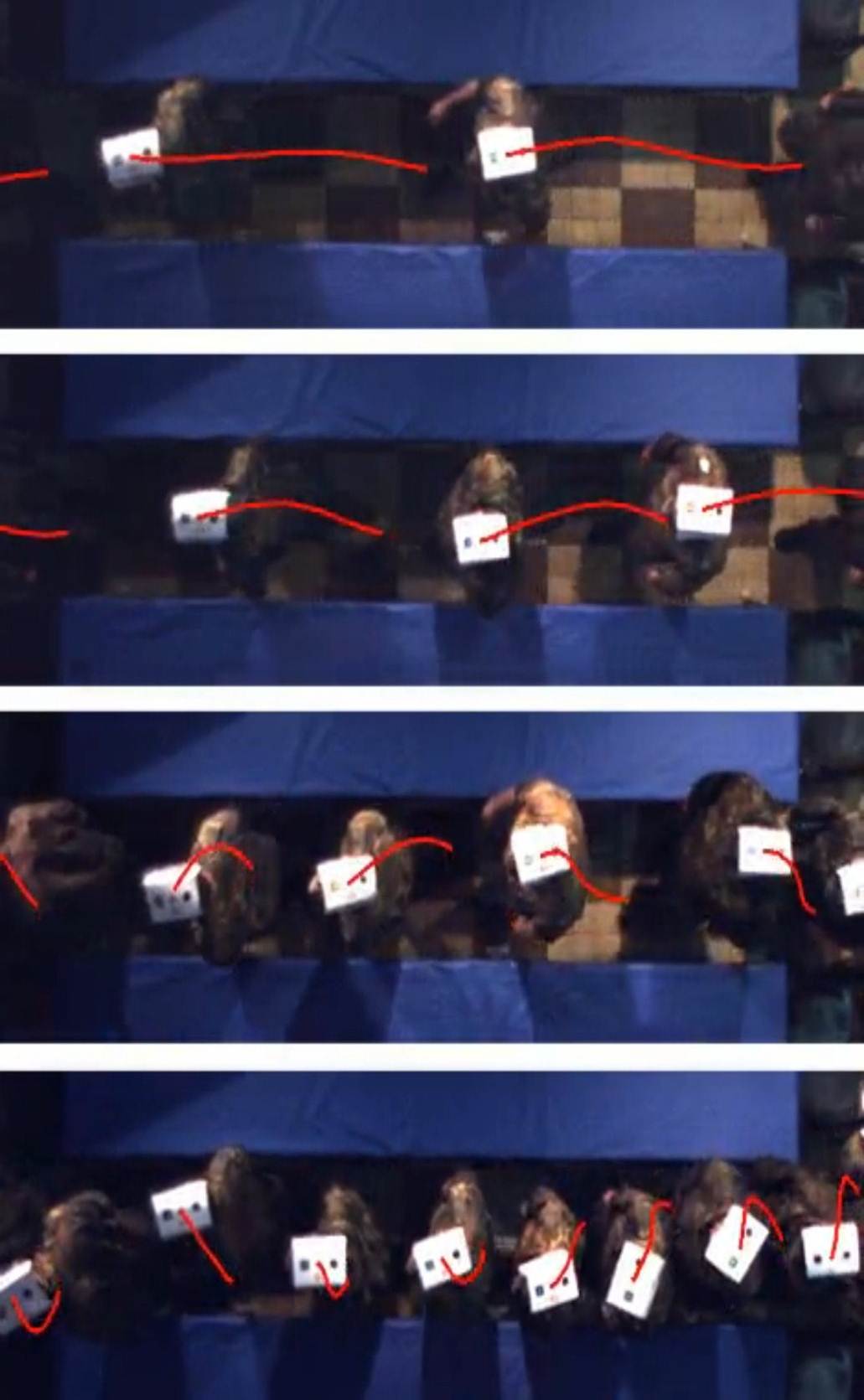}
\includegraphics[height=6 cm]{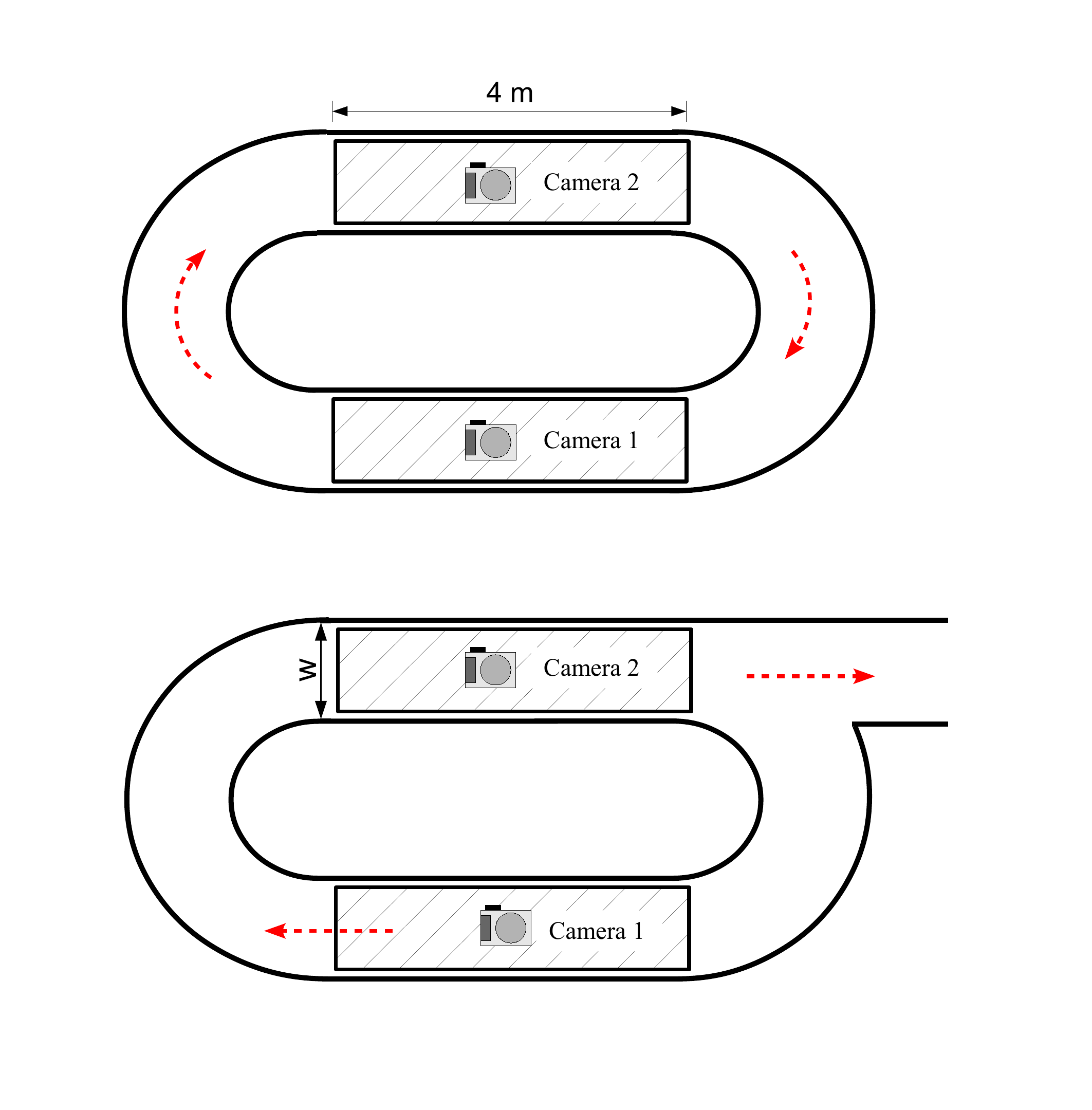}}
\caption{\label{fig-ped-exp} Snapshots and setup of the single file movement experiment. Two cameras are used to record the movement at different parts of the corridor. Camera 2 is close to the exit.}
\end{figure}

The experiment was performed in an oval corridor in the wardroom of Bergische Kaserne D\"usseldorf in Germany  in 2006. The circumference of the corridor was about $C = 26~m$. The participants were female and male soldiers. Pasteboards with high contrast markers were put on the head of each soldier for trajectory extraction. To get different ranges of global density in the corridor, 12 runs  were performed with the number of test persons N = 14, 17, 20, 22, 25, 28, 34, 39, 45, 56, 62, 70. This means that the global density ($\rho_g = N/C$) ranges from 0.54 to 2.69 $m^{-1}$ in this experiment.

Fig.~\ref{fig-ped-exp} shows the sketches and snapshots of the experiment. Two cameras (Camera 1 and 2) were used to record the pedestrian movement in the two 4 $m$ length areas with a frame rate of 25 $fps$.  At the beginning of each run, the pedestrians were arranged in the corridor uniformly. Then they were asked to move three to four rounds in the circuit in normal speed without overtaking. After that an exit near Camera 2 was made and pedestrian went outside the closed corridor.  As a result, the movements  both in closed and open situations were recorded from the experiment. Pedestrian trajectories were extracted from these video recordings  using the software PeTrack \cite{Boltes2010}. Detailed information about the experiment can be found in \cite{Seyfried2010b}.

\section{Time-space diagram}

\begin{figure}{
\includegraphics[width=0.45\textwidth]{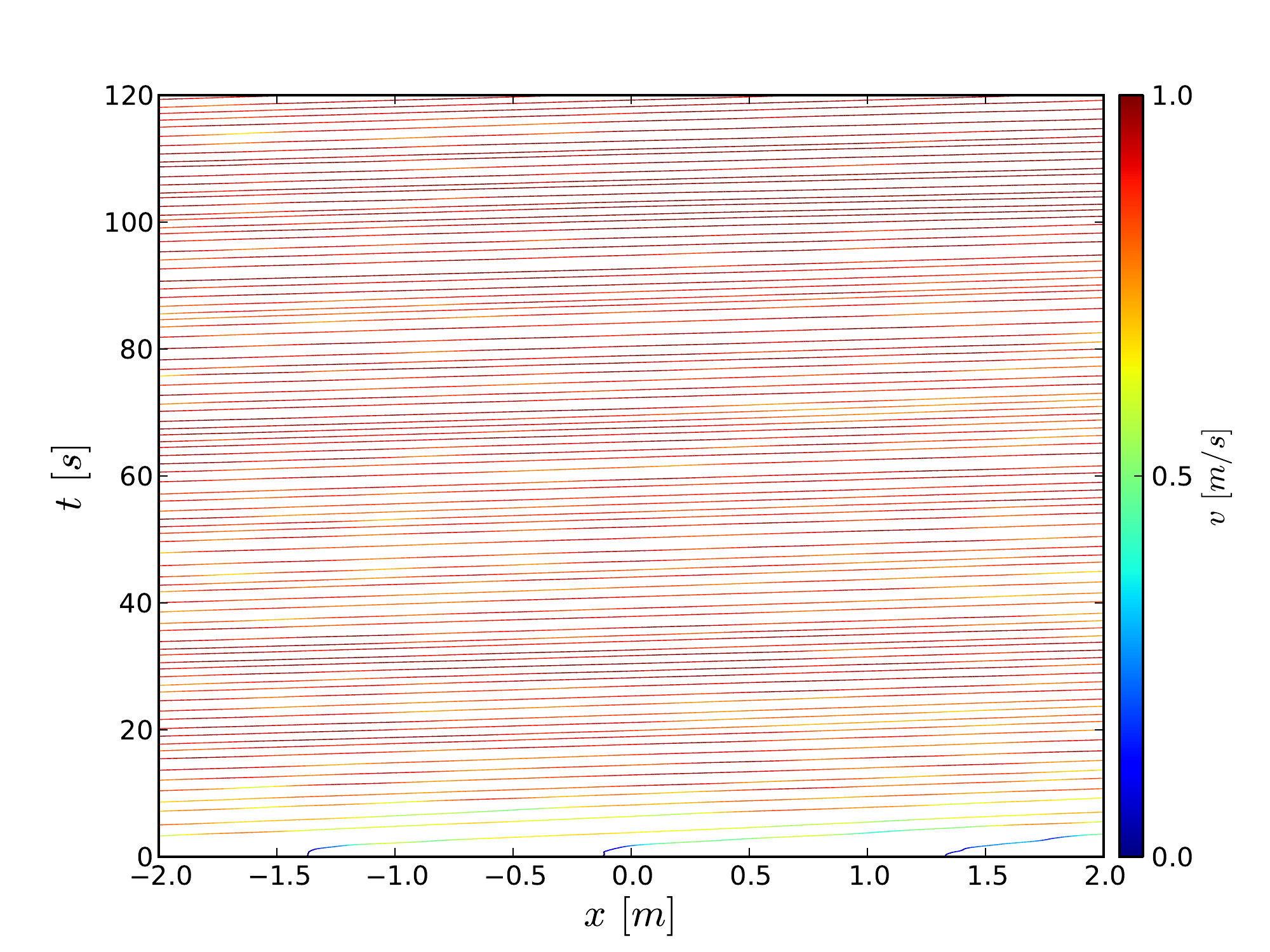}
\includegraphics[width=0.45\textwidth]{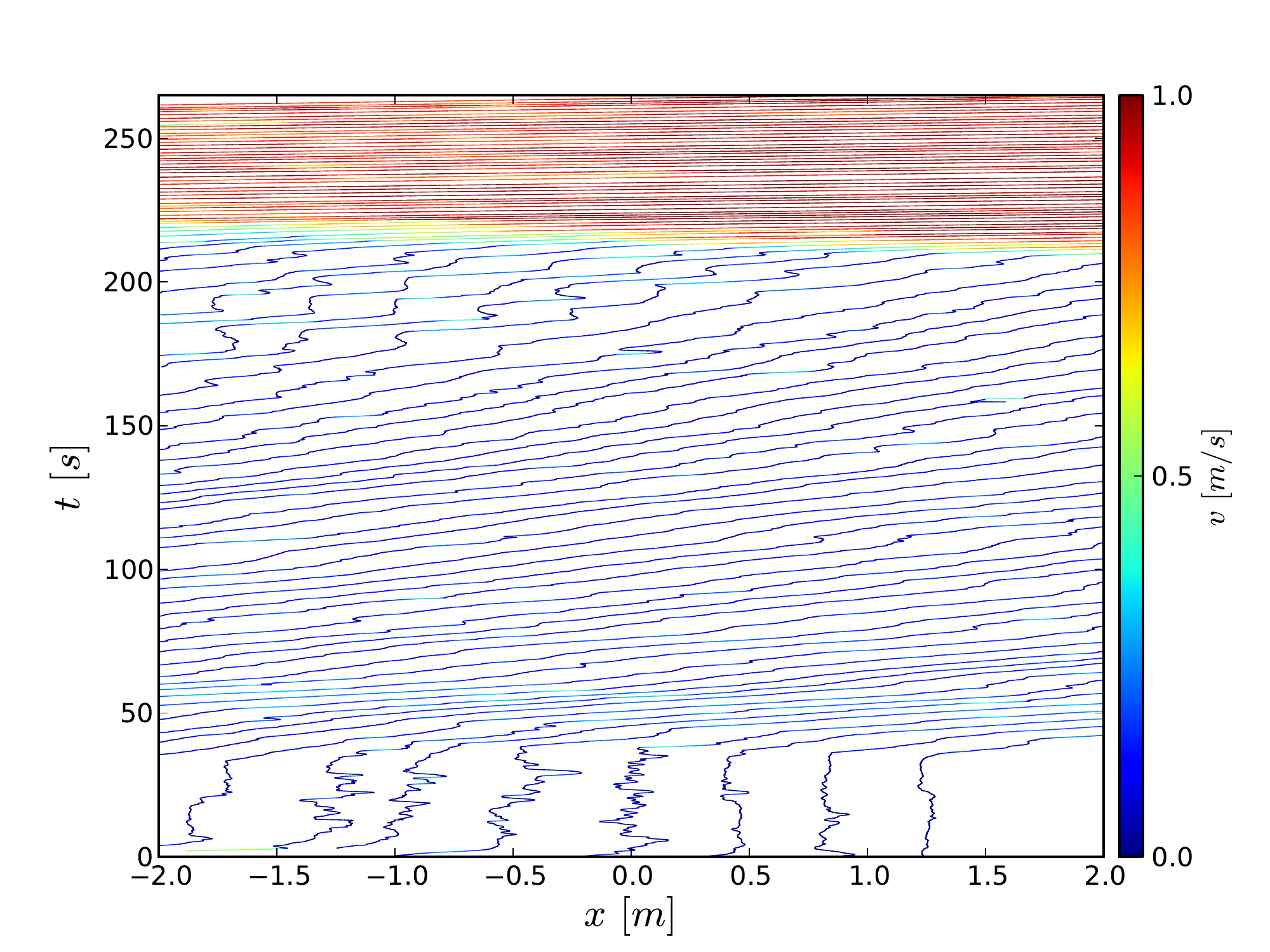}}
\caption{\label{fig-timespace-ped} Time-space diagram for four runs of experiment under both closed and open corridor conditions. There are 25 (left) and 62 (right) pedestrians in the corridor respectively.}
\end{figure}

Using the high precision trajectories extracted from the video recordings, some qualitative characteristics can be observed.
Fig.~\ref{fig-timespace-ped} shows the time-space diagrams for the runs with N = 25 and 62, which represents
the global density 0.96 and 2.38 $m^{-1}$. In these graphs, instantaneous velocities $v_i(t)$, which is calculated
according to the equation(\ref{eq2}) with $\Delta t^\prime = 10~frames$, of each pedestrian are also exhibited.

The data  in all these graphs come from the Camera 2 which captured the movement in both closed and open corridor. On one hand, different states of the pedestrian flow can be observed from these spatiotemporal diagrams. When there are 25 pedestrians in the corridor, the density is relatively low and the flow is at free flow state. With the increase of the number of pedestrians in  the corridor, the jamming state can be observed obviously and stop-and-go wave starts to appear. In the right figure for the run N = 62, the wave is so obvious and lasts for a long time. On the other hand, the influences of boundary condition on the flow are extremely evident especially at high density situation. When the exit is opened, the pedestrian flow transfer from congested state to free flow state quickly. The velocities of pedestrian increase from smaller than $0.05~m/s$ to more than $1~m/s$ rapidly.

\section{Analysis and results}

To gain a deeper understanding of the pedestrian dynamics, quantitative analysis is indispensable to uncover the underlying
dynamics that is not apparent through simple observations. In this section, the fundamental diagram, the basic relationship in
traffic engineering, will be investigated using more precise measurement method.

\subsection{Measurement method}

According to our previous studies on two dimensional pedestrian flow, the Voronoi-based method has the advantage
of high resolution and small fluctuation of measured densities compared to other methods \cite{Steffen2010a, zhang2011}.
At a macro level, we therefore use the concept of Voronoi method in this study.

\begin{figure}
\centering{
\includegraphics[width=0.9\textwidth]{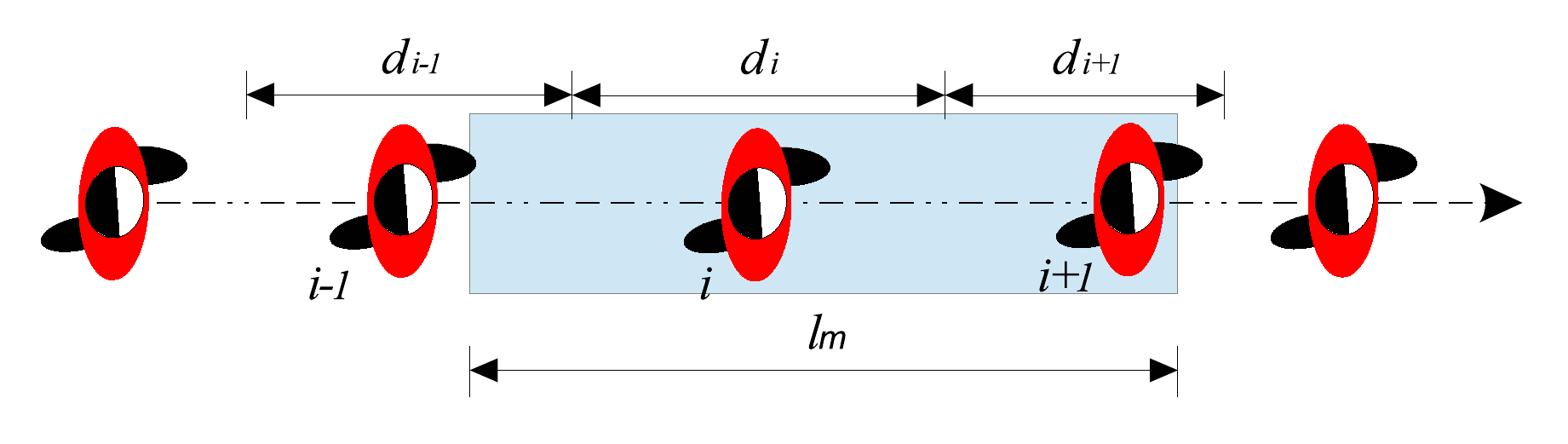}}
\caption{\label{fig-voro-sketch} Illustration of the Voronoi method in one dimensional space. The gray region shows the measurement area.}
\end{figure}

In two dimension space, the Voronoi space of a pedestrian is obtained based on the locations of his or her neighbors.
Similarly, in one dimensional case we calculate the Voronoi space $d_i$ of each pedestrian $i$ base on the positions his two
neighbors $i-1$ and $i+1$. Actually, the length of $d_i $  is half the distance between the two neighbors (see Fig.~\ref{fig-voro-sketch}).
In this method, a measurement area with the length of $l_m$ is selected to calculate the Voronoi density $ \rho (t)$ and
velocity $ v (t)$ at time $t$:

\begin{equation}\label{eq1}
 \rho (t)=\frac{\sum_{i=1}^{n}\Theta_i(t)}{l_m}\qquad
and \qquad  v (t)=\frac{\sum_{i=1}^{n}{\Theta_i(t) \cdot v_i(t)}}{l_m}
\end{equation}
where $n$ is the number of pedestrians whose Voronoi space includes the measurement area (assuming the overlapping length is $d_{oi}$ for
pedestrian $i$). $\Theta_i(t) = d_{oi}/d_i$ represents the contribution of pedestrian $i$ to the density of the measurement area. $v_i(t)$ is the
instantaneous velocity of pedestrian $i$ at time $t$. It is calculated by using his displacement in a small time interval $\Delta t^\prime$ around t,
that is
\begin{equation}\label{eq2}
v_i(t)=\frac{{x_i}(t+\Delta t^\prime/2)-{x_i}(t-\Delta t^\prime/2))}{\Delta t^\prime}
\end{equation}

In this study, a three meter measurement area from $x = -1.5~m$ to $x = 1.5~m$  is selected.  With such selection, it means that at most 7
pedestrians can exist in the area at the same time according to the average dimension of a standing pedestrian \cite{Fruin1971}.
The time interval $\Delta t^\prime = 0.4~s$ (corresponding to 10 frames) is used to calculate the velocity. We calculate the densities and velocities
every frame with a frame rate of $25~fps$.

\begin{figure}
\centering{
\includegraphics[width=0.45\textwidth]{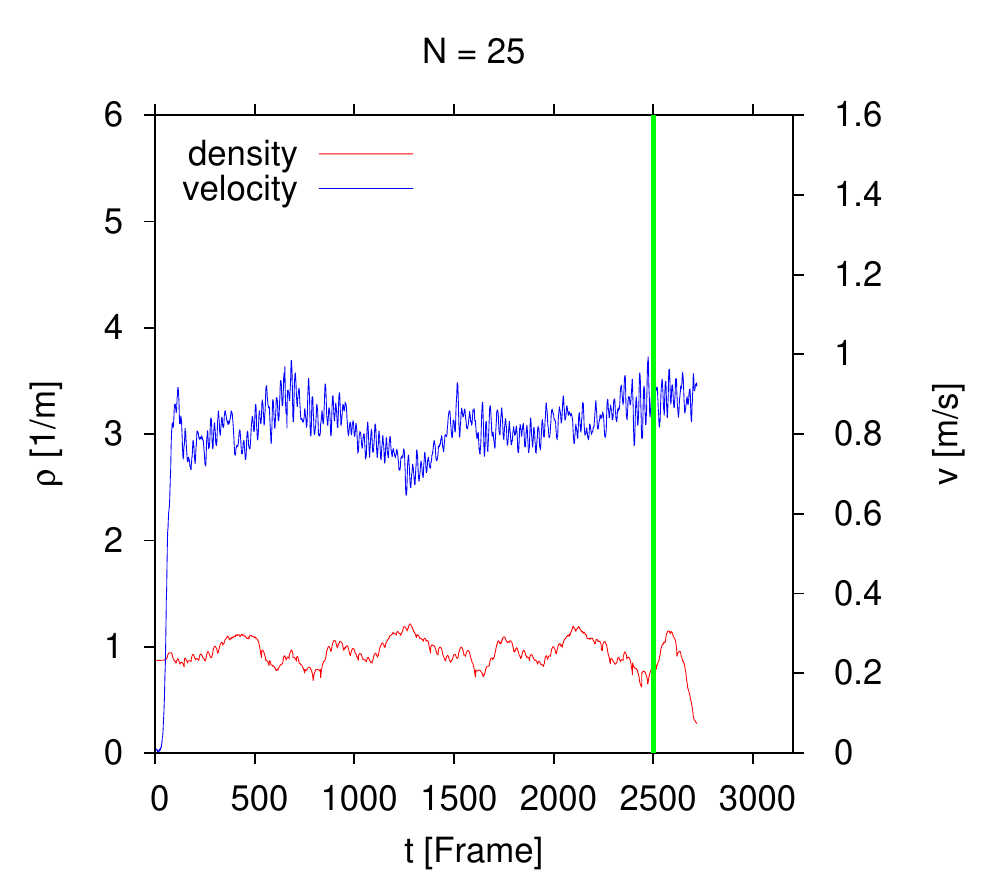}
\includegraphics[width=0.45\textwidth]{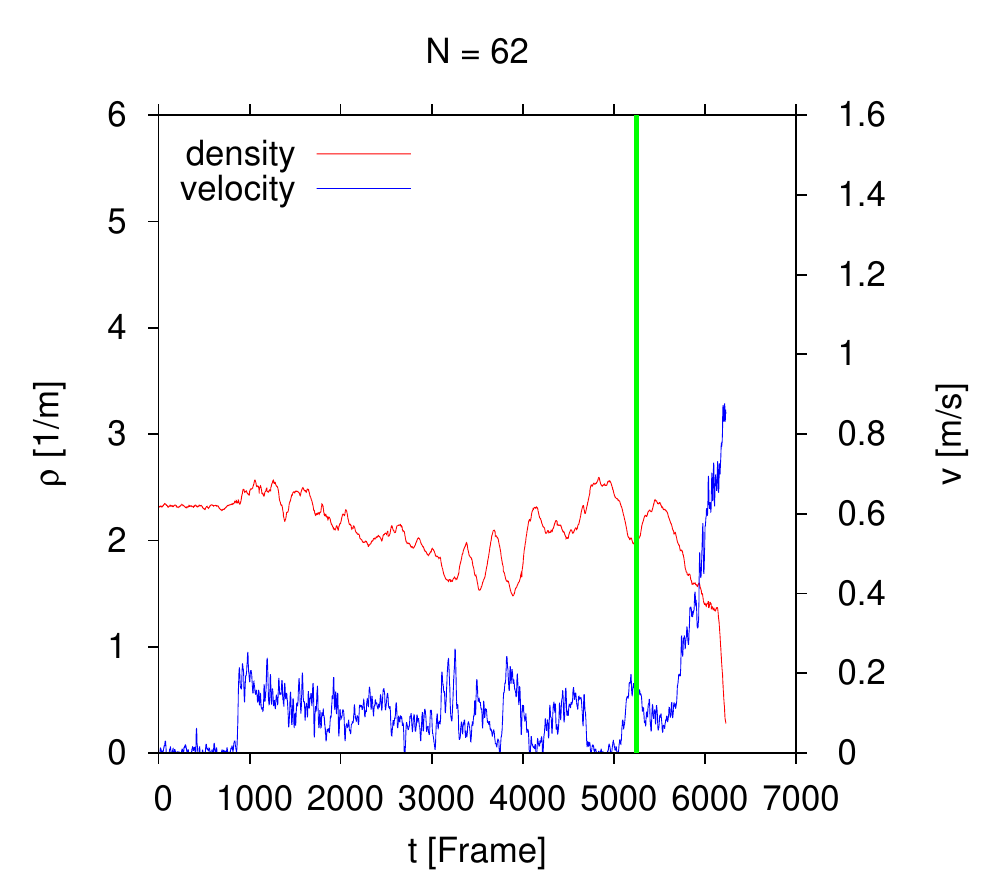}}
\caption{\label{fig-TS-camera1} Time series of the Voronoi density and velocity for two runs of the experiment. The green vertical line shows
the approximate time for opening the exit. The data here is from Camera 1. }
\end{figure}

\begin{figure}
\centering{
\includegraphics[width=0.45\textwidth]{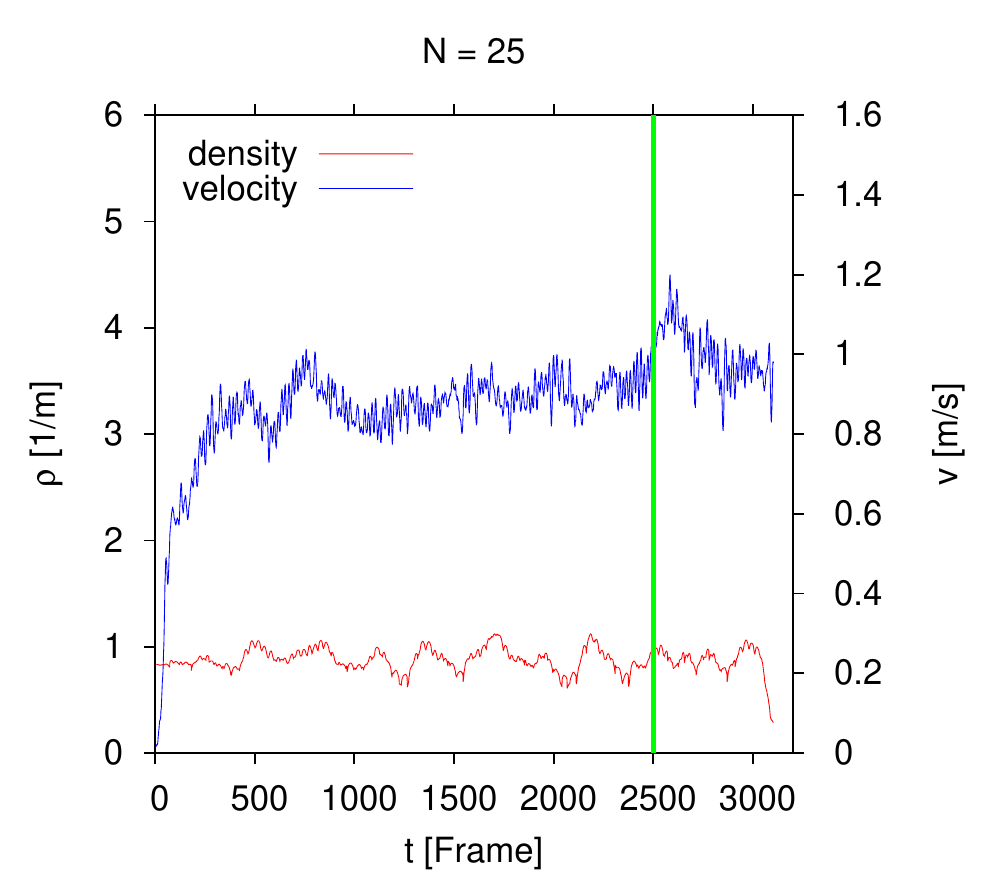}
\includegraphics[width=0.45\textwidth]{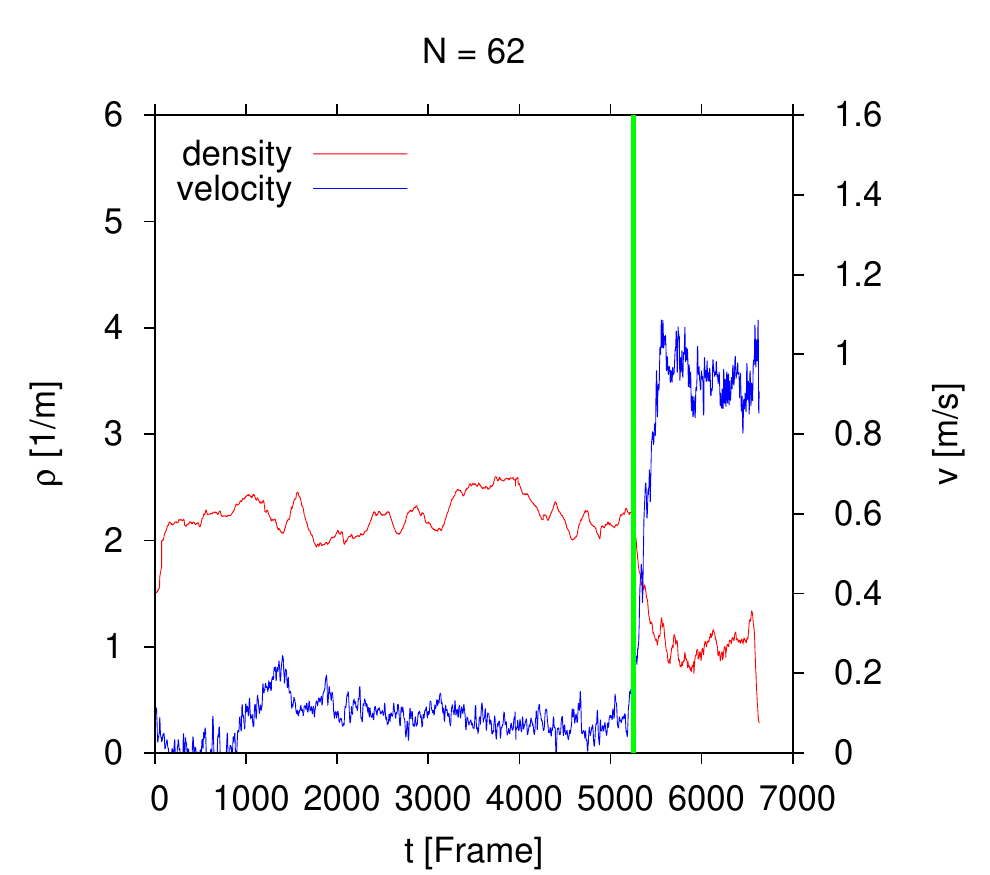}}
\caption{\label{fig-TS-camera2} Time series of the Voronoi density and velocity for two runs of the experiment. The green vertical line shows
the approximate time for opening the exit. The data here is from Camera 2.}
\end{figure}

To determine the fundamental diagram, it is necessary to use the data at stationary state which comes from the stable but not transient behavior of
the system. In this study, the stationary states are selected manually by analyzing the time series of density and velocity.  Fig.~\ref{fig-TS-camera1} and~\ref{fig-TS-camera2} show the time series for two runs of the experiment from the two cameras. The green
vertical lines in the graphs show the time for opening the exits. For the run with N = 25, the measured
values of the variables don't show marked differences no matter the corridor is open or closed and the flow is at free flow state. Conversely, the values change obviously when the flow is at jamming state at high density situation. For the run with N = 62 the effects are so clear that sudden drop of density and sharp rise of velocity (from 0.1 to 1 $m/s$) are observed. No large differences are found between the data from different cameras. However, the Camera 2 recorded relatively more data under open boundary condition and thus we mainly analyze the data from it in the following.

\subsection{Fundamental diagram}

\begin{figure}
\centering{
\includegraphics[width=0.45\textwidth]{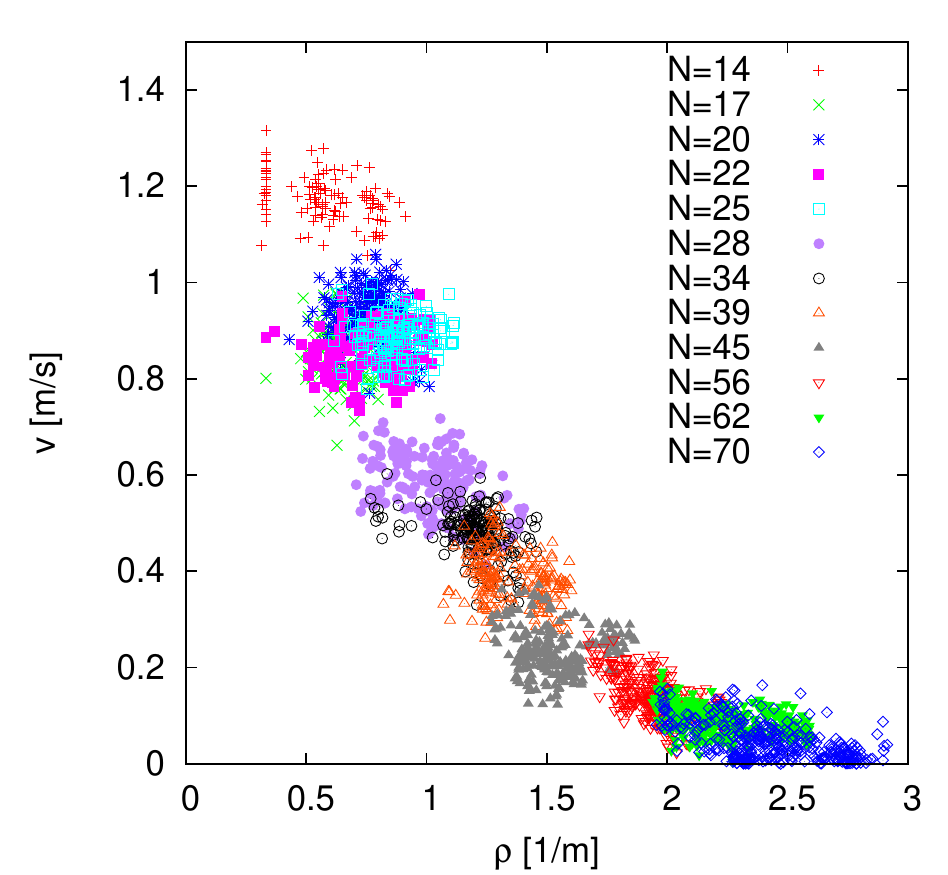}
\includegraphics[width=0.45\textwidth]{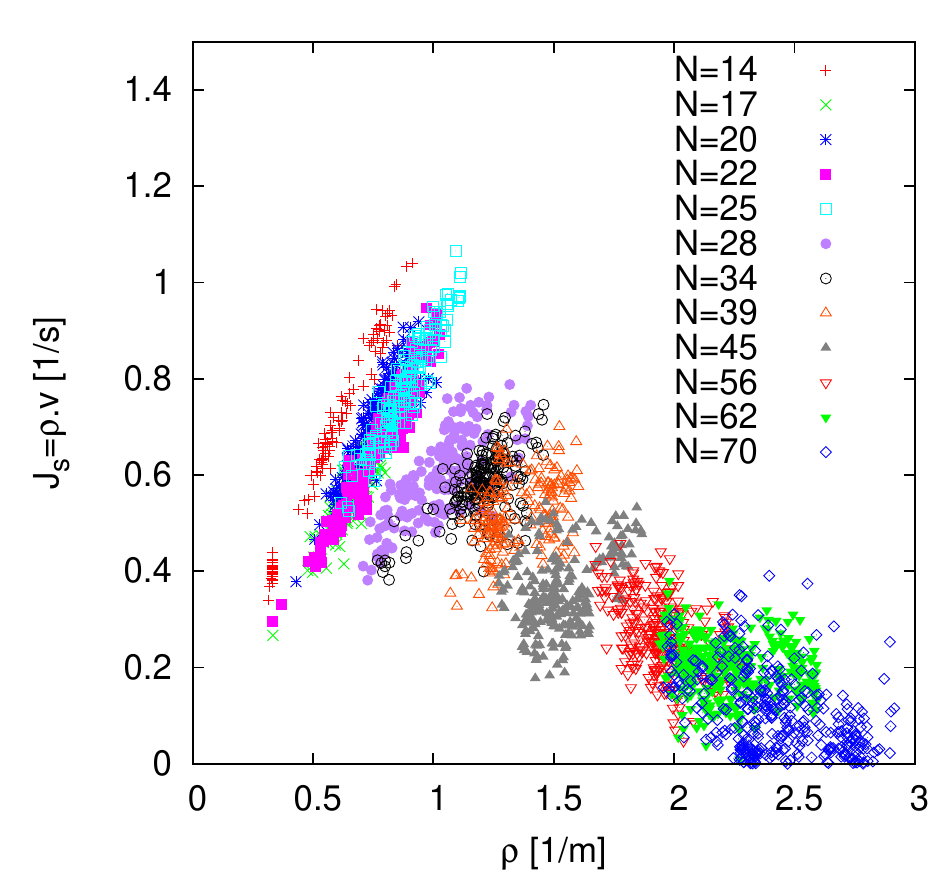}}
\caption{\label{fig-FD-Voro-closed} The fundamental diagram (density-velocity (left) and density-specific flow (right) relationships) of single file pedestrian movement in closed corridor.}
\end{figure}

Fig.~\ref{fig-FD-Voro-closed} shows the Voronoi-based fundamental diagram for the pedestrian movement under closed boundary condition. We plot the results from various runs in different symbols to see the changes in more detail. The free velocity is around 1.2 $m/s$ and is obtained from the run with N = 14. The maximum specific flow is about 1 $1/s$ around $\rho = 1~1/m$. For the four runs with N = 17, 20, 22 and 25, nearly no any difference can be observed from the data. The measure densities range from 0.5 to 1 $1/m$ and the velocities  are around 0.9 $m/s$. Whereas, after N = 28 the changes become apparent with the increase of the number of pedestrians inside the corridor. From this graph, it seems that the fundamental diagram could be divided into three parts for density $\rho$ belonging to [0, 1.0], [1.0, 1.7] and [1.7, 3.0]. These should be responding to the three states of pedestrian movement. When the density is smaller than $1.0~1/m$, it is free flow state and pedestrian can move smoothly. In the second part, it is congestion state and the specific flow begins to decrease. In this state, the decelerations of pedestrian can be observed sometimes but are not the main property of the movement. The data for $\rho > 1.7~1/m$ are mainly from the runs with  N =56, 62 and 70. In this state, the stop-and-go waves occupy the main time of pedestrian movement. Especially for the density around 1.7 $1/m$, the velocity seems to remain constant or increase in a small range. Reflected in the density-specific flow relationship, the specific flow has a transition around this density. Here another point worths noticing is that the change from free flow state to congestion state is not continuous. A gap seems exist in the density-velocity relationship around $v = 0.7~m/s$.

\begin{figure}
\centering{
\includegraphics[width=0.45\textwidth]{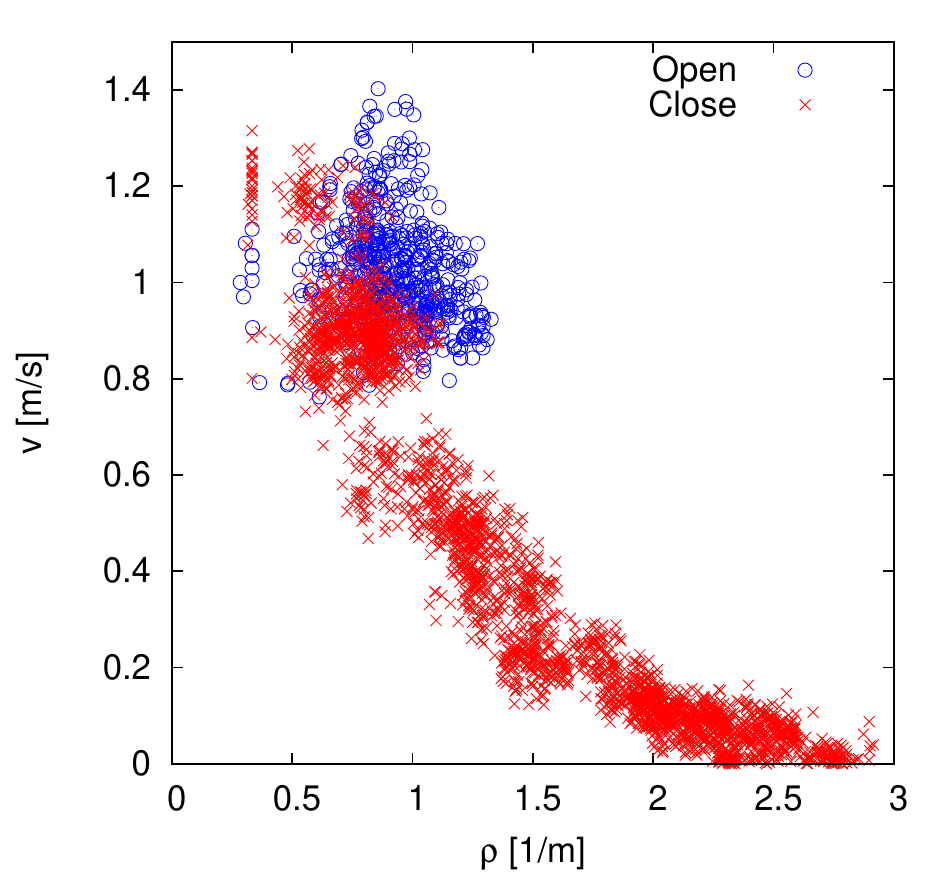}
\includegraphics[width=0.45\textwidth]{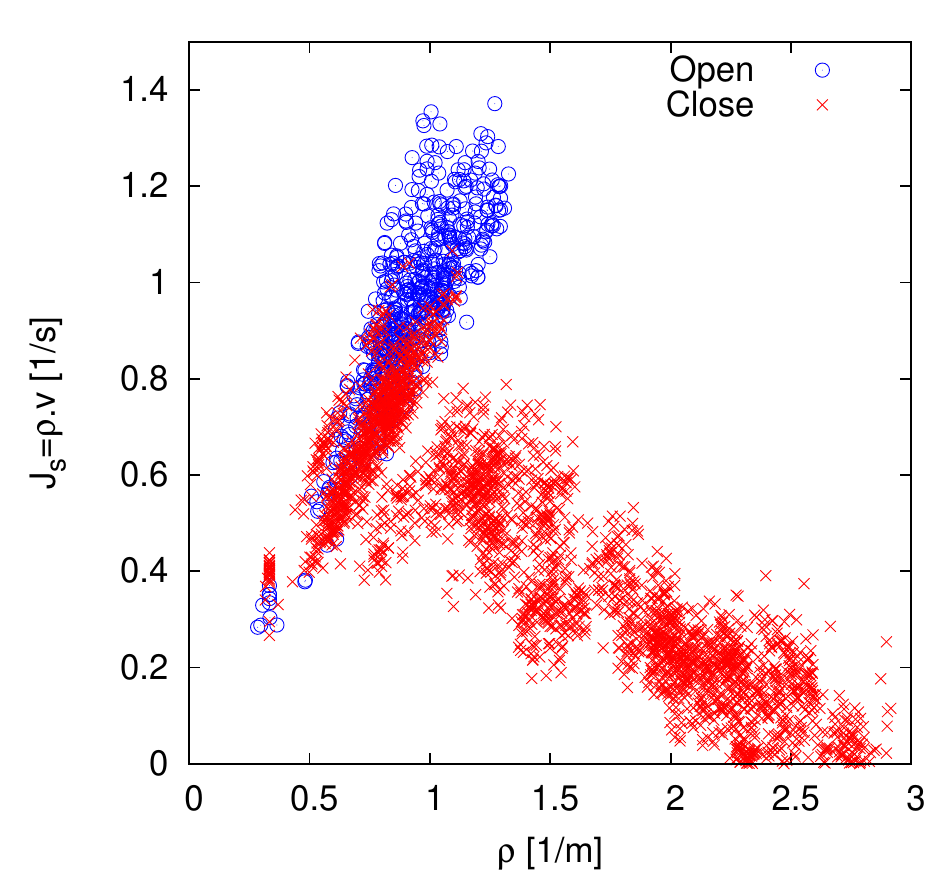}}
\caption{\label{fig-FD-Voro-comp} Comparison of the fundamental diagram (density-velocity (left) and density-specific flow (right) relationships) of single file pedestrian movement under different boundary conditions.}
\end{figure}

Next, we study the influence of boundary conditions on the fundamental diagram of single-file pedestrian flow. For open boundary condition only data at free flow state are observed in this study. The maximal density obtained from this situation is less than 1.3 $1/m$, as shown in Fig.~\ref{fig-FD-Voro-comp}. Under the same density situation, the velocities are obviously higher in open corridor case. Correspondingly the maximal specific flow increase from 1.0 $1/s$ at closed boundary to 1.3 $1/s$ at open boundary.

\section{summary}

In this paper, we study the properties of single file pedestrian movement in oval corridor by means of an experiment performed under laboratory conditions.
We investigate the influence of boundary condition on the flow both quantitatively and qualitatively.

From the time-space diagram, the characteristics of different flow states can be observed qualitatively. The influence of boundary on the flow is extremely high at jamming state. The flow can change from jamming flow to free flow quickly. The velocities of pedestrian can increase from smaller than 0.05 $m/s$ to more than 1 $m/s$ rapidly.

From the analysis both at micro and macro level, it is found that the boundary condition do have influence on the fundamental diagram of the single file pedestrian flow. At the same density situation, the velocities of pedestrians are higher under open boundary condition. The measured maximal specific flow is larger than 1.3 $1/s$, which is obviously larger than that under closed boundary condition.However, the maximal density obtained at open boundary condition is less than 1.3 $1/m$. Further empirical data in this situation is still necessary to study the influence at high density situation.

\end{document}